\documentstyle[epsf,twocolumn]{article}

\newtheorem{definition}{Definition}

\title{Generalized Singular Spectrum Time Series Analysis}

\author{Martin Nilsson \\ 
	EES-6, MS T003 \\
    	Los Alamos National Laboratory\\
    	{\tt nilsson@lanl.gov}}

\begin{document}
\maketitle

\begin{abstract}
{\bf This paper is a study of continuous time Singular Spectrum Analysis (SSA). 
We show that the principal eigenfunctions are
solutions to a set of linear ODEs with constant coefficients. 
We also introduce a natural generalization of SSA, constructed using local (Lie-)
transformation groups. The time translations used in standard SSA is a special case. The 
eigenfunctions then satisfy a simple type of linear ODE with time dependent
coefficient, determined by the infinitesimal generator of the transformation group.
Finally, more general one parameter mappings are considered.}
\end{abstract}

Singular Spectrum Analysis (SSA) is a relatively recent method for nonlinear
time series analysis. The original idea behind SSA was first presented by 
Broomhead and King~\cite{Broomhead}, in the context of time series embedding.  
During the last decade this technique 
has been very successful and has become a standard tool in many 
different scientific fields, such as climatic~\cite{Vautard}, meteorological~\cite{Ghil}, 
and astronomical~\cite{Vardi} time series analysis.
For introductions to the SSA technique, see e.g.,~\cite{Elsner,Golyandina}.

In practical applications, a time series is a result of a
sampled measurement, and is therefore discrete. 
This paper is a theoretical study of SSA, and  it is
therefore more natural to consider the general case of continuous time. 
We start the paper by a (semi) formal expansion of the SSA procedure to 
continuous time.

Let $f (t)$ be a function representing a continuous time signal on an interval
$\Omega _T = [0,T]$. We assume that $f (t) \in {\cal L}_2 (\Omega _T )$, 
where ${\cal L}_2 (\Omega _T)$ is a Hilbert space with an inner product defined as 
$(f , g) _{\Omega _T} = \int _{\Omega _T} dt f (t) \overline{g (t)}$  
and a norm $ \| f \| _{\Omega} = \sqrt{ (f , f )_{\Omega}}$. We define a {\em trajectory function} 
$X (\xi , t)$
\begin{eqnarray}
	X (\xi,t) & = & f( \xi + t ) 
\label{trajectory}
\end{eqnarray}
where $t \in \Omega _W  = [0,W]$ and $\xi \in \Omega_{T - W} = [0,T-W]$.  
The parameter 
$W$ is fixed and referred to as the {\em window length}. By construction $ X (\xi,t) \in 
{\cal L} _2 (\Omega _{T-W} ) \times {\cal L} _2 (\Omega _{W} )$ and the
norm is defined as $\| X \| ^2 = \int  _{\Omega _{T-W} } d \xi \int _{\Omega _W} d t
X (\xi , t ) \overline{ X (\xi , t ) }$.

A Schmidt decomposition (the continuous equivalent to a Singular Value Decomposition
of a matrix) of the trajectory function is defied as 
\begin{eqnarray}
	X ( \xi , t) & = & \sum _{k=1} ^d \sqrt{\lambda _k } v _k ( \xi ) u_k ( t )
\label{decompose}
\end{eqnarray}
where $d$ is referred to as the rank (which is often infinite). 
All partial sums, $k=1$ to $\tilde{d}$, are optimal 
$\tilde{d}$-rank ${\cal L}_2$-approximations of $X ( \xi ,t )$, i.e.,
the functions $v_k (\xi )$ and $u_k (t)$ fulfills
\begin{eqnarray}
	\min _{ {\bf u} , {\bf v} } 
		\| X (\xi , t) - \sum _{k=1}^{\tilde{d}} \sqrt{ \lambda _k} 
		v _k (\xi ) u _k (t) \| 
\label{minimization}
\end{eqnarray}
Without lost of generality, the functions $v_k (\xi )$ and $u_k (t)$ can be assumed
to be normalized, $\| u_k \| = \| v_k \| = 1$. Eq.~\ref{minimization}
implies
\begin{eqnarray}
	v _k (\xi) & = & \frac{1}{\sqrt{\lambda _k}} \int _{\Omega _{W}} d t
		X (\xi,t) \overline{u _k (t)} \nonumber \\
	u _k (t) & = & \frac{1}{\sqrt{\lambda _k}} \int _{\Omega_{T-W}} 
		\! \! \! \! \! \! \! d \xi X (\xi,t) \overline{v _k (\xi)} 
\label{opt_cond}
\end{eqnarray}
together with the orthogonality conditions $\left( v _k , v_l \right) _{\Omega _{T-W}} = 
\delta _{k l}$ and  $\left( u _k , u_l \right) _{\Omega _T} = 
\delta _{k l}$. The two relations in Eq.~\ref{opt_cond} can be combined into
eigenvalue problems for the functions $u _k (t)$ respective  $v _k (\xi )$
\begin{eqnarray}
	\int _{\Omega _W} d t^{\prime} S (t, t^{\prime}) u _k (t^{\prime}) & = & \lambda _k u _k (t) \nonumber \\
	 \int _{\Omega _{T-W}}  \! \!    d \xi ^{\prime} R (\xi , \xi ^{\prime}) 
		v_k (\xi ^{\prime}) & = & \lambda _k v_k (\xi )
\label{eigen_u_v}
\end{eqnarray}
where the {\em $t$-covariance function} $S (t,t^{\prime})$ and
the {\em $\xi$-covariance function} $R (\xi , \xi ^{\prime})$ are defined as
\begin{eqnarray}
	R (\xi,\xi ^{\prime}) & = & \int _{\Omega _W}  d t X (\xi ,t) 
	\overline{X (\xi ^{\prime} , t)} \nonumber \\
	S (t, t^{\prime}) & = & \int _{\Omega_{T-W}} \! \! \! \!  \! \! d \xi X (\xi , t) 
	\overline{X (\xi , t^{\prime})} \nonumber 
\end{eqnarray}
From Eq.~\ref{eigen_u_v} it is clear that $v_k (\xi )$ and $u _k (t)$ are 
eigenfunctions to two compact, linear and symmetric integral operators $R$ and $S$
with kernels  $R (\xi , \xi ^{\prime} )$ and
$S(t,t^{\prime})$. The spectral theorem
then guarantees $v_k (\xi)$ and $u_k (t)$ to form complete orthogonal
bases in ${\cal L}_2 (\Omega _{T-W}) / {\cal N} (R)$ respective ${\cal L}_2 (\Omega _T)
/ {\cal N} (S)$ (where ${\cal N} ( \cdot )$ denotes a null-space). The eigenvalues
$\lambda _k$ are real and non-negative, and measure the variance of $X (\xi , t)$ in the ``direction''
defined by $v _k (\xi )$ and $u _k (t)$. Note the close correspondence to principal 
component analysis.

By construction, SSA decomposes the original time series 
into orthogonal components. Usually, the components also represent intuitive 
contributions to the time series, such as a trend, various oscillation 
modes and noise.

We start our analysis of the SSA technique by investigating signal separation
into orthogonal subspaces.  From now on, due to space limitations, 
we will focus our attention to the ``right eigenmodes'', $u _k (t)$. By symmetry, 
equivalent results are valid for the ``left eigenmodes'', $v _k ( \xi )$. 
Assume that the time series can be decomposed as $f(t) = \sum _i f_i (t)$, where
\begin{eqnarray}
	\int _{\Omega _{T-W}} \! \! \! \! \! \! \! \! \! \! d \xi f_i ( t + \xi ) 
		f_j ( t^{\prime} + \xi ) & = & C_1 \cdot \delta _{i j} \nonumber \\
	\int _{\Omega _{W}} \! \!  d t f_i ( t + \xi ) 
		f_j ( t + \xi ^{\prime} ) & = & C_2 \cdot \delta _{i j} 
\label{f_ort}
\end{eqnarray}
$\forall t , t^{\prime} \in \Omega _W , \xi, \xi ^{\prime} \in \Omega _{T-W}$. 
Then the t-covariance kernel
also decompose as $S = \sum _i S _i$, where $ S _i $ corresponds to $f_i $. Since
$\int _{\Omega _W} d t S _i \left( t , t^{\prime} \right) S _j \left( t ^{\prime} , 
t^{\prime \prime} \right) = 0$ if $i \neq j$, it follows that $S _i u = \lambda u$
implies $u \in {\cal N} \left( S _j \right)$, $j \neq i, \lambda \neq 0$. An 
eigenfunction of $S_i$ is therefore also an eigenfunction of $S$, i.e.,
$S _i u = \lambda u$ implies $S u = \lambda u$. Since $S$ a symmetric operator,
$u_i \perp u _j$ when $\lambda _i \neq \lambda _j$, which guarantees that a
Schmidt decomposition is unique up to a rotation of eigenfunctions
with identical singular value. It follows that if $f(t) =  \sum _i f_i (t)$,
Eq.~\ref{f_ort} is fulfilled and $f_i (t)$ and $f_j (t)$ have disjoint spectra,
then the SSA decomposition in Eq.~\ref{decompose} is a direct sum of the
decompositions of the individual time series. In this case $f_i (t)$ and $f_j (t)$ 
are called {\em strongly separable}. If the spectra are not disjoint, but
Eq.~\ref{f_ort} is fulfilled, $f_i (t)$ and $f_j (t)$ are called {\em weakly
separable} (using the same notation as in~\cite{Golyandina}).

An important implication of the above analysis is that, for periodic functions, 
when the total time frame $T$ and the time window $W$ is chosen such 
that $T-W = n \tau$ and $W =  n^{\prime} \tau$ (where $\tau$ is the
period and $n , n^{\prime} \in {\bf Z}^+$), the singular spectrum $\sqrt{\lambda _k}$ is
identical to the Fourier coefficients of $f (t)$. Furthermore, if the
Fourier coefficients are distinct, then the eigenfunctions $u _k (t)$ 
are identical to the basis functions in the Fourier expansion. Similar results
hold asymptotically in the infinite time frame limit.

To understand the SSA procedure, it is essential to further analyze the 
characteristics of the orthogonal eigenfunctions $u_k (t)$. We start by the 
following trivial observation:
\begin{eqnarray}
	f \left( \left( \xi + \xi ^{\prime} \right) + t \right) 
	& = & f \left( \xi + \left( \xi ^{\prime} + t  \right) \right).
\label{trivial}
\end{eqnarray}
Using this relation in Eq.~\ref{decompose} gives
\begin{eqnarray}
	 \sum _{k=1} ^d \sqrt{\lambda _k } v _k ( \xi ) u_k ( \xi ^{\prime} + t )   & = & \nonumber \\
	 \sum _{k=1} ^d \sqrt{\lambda _k } v _k ( \xi + \xi ^{\prime}) u_k ( t ) .  & & 
\nonumber
\end{eqnarray}
Apply the projection operator $\int _{\Omega _{T-W}} \!\!\!  d 
\xi v_l ( \xi )$, use the 
orthogonality of $v_k (\xi )$, subtract $u _l (t)$, divide by $\xi ^{\prime }$, and 
finally let $\xi ^{\prime} \rightarrow 0$. Technically, a problem appears when integrating 
$v_k (\xi + \xi ^{\prime})$ over $\xi \in \Omega _{T-W}$. The function is not defined
when $\xi + \xi^{\prime} \notin \Omega _{T-W}$.
However, since $\xi ^{\prime} \rightarrow 0$, using any smooth continuation of $v_k$
gives equivalent results. Alternatively, Eq.~\ref{trivial} could be replaced by 
$\partial _t f (\xi + t ) = \partial _{\xi} f (\xi + t )$. Either way, we find:
\begin{eqnarray}
	\partial _t u _k (t) & = & \sum _{l=1}^d A _{k l} u _l (t) ,
\label{u_standard}
\end{eqnarray}
where 
\begin{eqnarray}
	A _{k l} & = & \sqrt{ \frac{\lambda _l}{\lambda _k}} \int _{\Omega _{T-W}} \! \! \! \! \! 
	\! \! \! \! \!   d \xi v _k ( \xi ) \partial _{\xi} v _l  ( \xi ) 
\nonumber
\end{eqnarray} 
does not depend on $t$. Note the connection between $A$ and the cross-correlation
function. Projecting Eq.~\ref{u_standard} onto $u _k (t)$ also gives:
\begin{eqnarray}
	A _{k l} & = & \int _{\Omega _W} d t u_l (t) \partial _t u _k (t)  \nonumber
\label{u_v_symm}
\end{eqnarray}
which clearly shows the symmetric relation between the left and right principal 
eigenfunctions.

From Eq.~\ref{u_standard} it is clear that polynomial, exponential and harmonic functions
show especially simple (finite) spectra during SSA. In the more general case,
the SSA procedure decomposes the time series into an optimal 
(infinite) linear combination of polynomial, exponential and harmonic functions.
This has previously been discovered by others, but the approach 
taken in this paper is quite different and arguably more straight forward. For a discussion
on related work see~\cite{Vardi,Golyandina}.

We now show a connection between the matrix ${\bf A}$ and an underlying
dynamical system from which the time series is generated.
Let ${\bf f} (t)$ be a solution to a linear system of differential equations
\begin{eqnarray}
	\partial _t {\bf f} (t) & = & {\bf M} {\bf f} (t),
\nonumber
\end{eqnarray}
and let a scalar time series $f (t)$ be defined by some linear projection of
${\bf f} (t)$, $f = {\bf p}^T {\bf f} $, where ${\bf p}$ is non-degenerate in the sense 
that all oscillation modes that appears in ${\bf f} (t)$ are also present in $f (t)$. 
We shall now prove that ${\bf M}$ and 
${\bf A}$ have the same spectra. The vector function ${\bf f} (t)$ can be
decomposed as:
\begin{eqnarray}
	{\bf f} (t) & = & \sum _{k=1}^{d} \sqrt{\widetilde{\lambda} _k} {\bf w} _k 
		\widetilde{u} _k (t) ,
\label{fat_f}
\end{eqnarray}
where $t$ is restricted to the interval $\Omega _W$. Since ${\bf f} ( \xi + t) = 
\exp ( {\bf M} \xi ) {\bf f} (t)$, we have
\begin{eqnarray}
	f (\xi +t ) = \sum _{k=1}^{d} \sqrt{\widetilde{\lambda} _k} \left( {\bf p}^T 
	\exp \left\{ {\bf M} \xi \right\} {\bf w}_k \right) \widetilde{u}_k (t) . \nonumber
\end{eqnarray}
Comparing this expression to Eq.~\ref{decompose} and use the uniqueness of the SVD expansion,
gives $v _k ( \xi) = C_k \cdot  \sum _l R _{k l} \left({\bf p}^T \exp \left\{ {\bf M} \xi \right\} 
{\bf w}_l \right)$, where $C_k$ is a normalization constant and ${\bf R}$ is some rotation matrix 
which rotates elements within the equivalence classes defined by identical singular values. 
This further shows that $\widetilde{{\bf u}} (t) = {\bf R} {\bf u} (t)$, and therefore
the time derivative of Eq.~\ref{fat_f} gives:
\begin{eqnarray}
	\sum _{k  l} \sqrt{\widetilde{\lambda} _k} {\bf M} {\bf w} _k R _{k l} u _l (t) 
		& = & \nonumber \\
	\sum _{k  l m} \sqrt{\widetilde{\lambda} _k} {\bf w} _k R_{k l} A _{l m} u _m (t) . &&
\nonumber
\end{eqnarray}
Using the orthogonality of $u_k(t)$ and ${\bf w} _k$, gives
\begin{eqnarray}
	{\bf A}  & = &  {\bf R}^T \widetilde{\Lambda} ^{-1} {\bf W}  ^{T} {\bf M}  {\bf W} 
		\widetilde{\Lambda} {\bf R},
\label{A_M}
\end{eqnarray}
where we use the matrix notation: $\widetilde{\Lambda } = \mbox{diag} 
\left( \sqrt{\widetilde{\lambda} _1}, \sqrt{\widetilde{\lambda} _2}, \dots \right)$, 
${\bf W} = \left( {\bf w}_1 , {\bf w}_2 , \dots \right)$. 
Since $ \left( {\bf W} \widetilde{\Lambda} {\bf R} \right) \left( {\bf R}^T \widetilde{\Lambda} ^{-1}  {\bf W}^T
\right) = {\bf 1}$, Eq.~\ref{A_M}
implies that ${\bf A}$ and ${\bf M}$ have identical spectra.

Note that this line of argument is similar to the local linear analysis of dynamical
systems, used as part of the proof delay coordinate embedding theorems~\cite{Takens}.
It also shows a straight forward connection between PCA of a set of time series from 
a system of ODEs, and a SSA analysis of a projection from the system.

Though natural in the discrete case, the definition of the trajectory 
function as $X ( \xi , t) = f ( \xi + t )$ is somewhat arbitrary
in our analysis. Why not $X ( \xi , t) = f ( \xi t )$, for example?
We can use this arbitrariness both to generalize SSA and
to gain better theoretical understanding of the procedure. Recall the 
definition of a (local) transformation group (see e.g.,~\cite{Olver}:
\begin{definition}
A transformation group is a continuous Lie group $G$ and a
set  $M \subset {\bf R}^n$ along with a smooth map
$\Psi : G \times M \rightarrow M$ which satisfies, for every
$g,h \in G, x \in M$,
\begin{eqnarray}
	\Psi \left( g , \Psi \left( h , x \right) \right) & = &
	\Psi \left( g \circ h , x \right) ,
\label{trans_prop}
\end{eqnarray}
together with the existence of an identity element $e$ and an inverse
$g^{-1}$ for all $g \in G$.
\label{trans}
\end{definition}
In this paper, the elements of the transformation group is spanned
by a parameter $\xi$ (we have a one-parameter group). We will sometimes 
use the  compact notation:
\begin{eqnarray}
	\Psi \left( g \left( \xi \right) , x \right) & \equiv & \Psi _{\xi} x .
\nonumber
\end{eqnarray}
We also chose the parametrization such that $g(0) = e$.

A one-parameter transformation group defines a vector field ${\bf V}
= \sum _i \zeta _i ( x_i ) \partial _{x_i}$: 
\begin{eqnarray}
	\left. {\bf V} \right| _x & = & \left. \frac{d}{d \xi} \right| _{\xi =0}
		\Psi ( \xi , x ) , \nonumber
\end{eqnarray}
which formally is ``solved'' as
\begin{eqnarray}
	\Psi ( \xi , x ) & = & \exp \left\{ \xi {\bf V} \right\} x . \nonumber
\end{eqnarray} 
The vector field is therefore called the {\em infinitesimal generator} of the
one-parameter transformation group. Note that the exponential map also implies 
$g ( \xi ) \circ g ( \xi ^{\prime} ) = g ( \xi + \xi ^{\prime} )$.

The transformation group also generates orbits through every point $x \in M$, 
defined as $\phi (\xi ) = \Psi ( \xi , x )$. The orbits are solutions
to a system of ordinary differential equations:
\begin{eqnarray}
	\partial _{ \xi } \phi _i (\xi )  & = & \zeta _i ( \phi  (\xi ) ) \nonumber \\
	\phi  ( 0 ) & = & x   ,
\label{orbit}
\end{eqnarray}
where the explicit representation of the vector field is used. Eq.~\ref{orbit}
can be used to derive explicit expressions for the transformation
group corresponding to a vector field.

Using this framework, the SSA procedure may be generalized in the following way.
Consider a trajectory  function $X_G$, constructed from a one-parameter transformation group
$\Psi _{\xi}$ acting on $M = \Omega _W$, with the continuous Lie group $G$:
\begin{eqnarray}
	X _G ( \xi , t ) & = & f \left( \Psi _{\xi } t \right) ,
\label{gen_traj}
\end{eqnarray}
where $\Psi : \Omega _{T-W} \times \Omega _W \rightarrow \Omega _T$.
Eq~\ref{trans_prop} provides a relation equivalent to 
Eq.~\ref{trivial}:
\begin{eqnarray}
	f \left( \Psi _{\xi} \left( \Psi _{\xi ^{\prime}} t \right) \right) & = &
	f \left( \Psi _{\xi + \xi^{\prime} } t  \right) ,
\label{gen_trivial}
\end{eqnarray}
Eq.~\ref{gen_trivial} can be used to find a system of differential 
equations, equivalent to Eq.~\ref{u_standard}, in terms of the infinitesimal generator 
of the transformation group:
\begin{eqnarray}
	{\cal L} \left[ u _k (t) \right] & = & \sum _{l=1}^d A_{k l} u_l (t),
\label{u_general}
\end{eqnarray}
where ${\cal L} \left[ \cdot \right]$ expresses a Lie derivative with respect 
to the 
transformation group. Since $u_k (t)$ is a scalar function (the index $k$ is fixed), 
${\cal L} \left[ u _k \right] = \zeta (t) 
\partial _t u _k (t)$, where $ \zeta (t) \partial _t$ is the vector field 
generating the one-parameter transformation group. The matrix ${\bf A}$ is now defined as
\begin{eqnarray}
	A _{k l} & = & \sqrt{ \frac{\lambda _l}{\lambda _k}} \int _{\Omega _{T-W}} \! \! \! \! 
	\! \! \! \! \! \!   d \xi v _k ( \xi ) \partial _{\xi} v _l ( \xi ) .
\nonumber
\end{eqnarray}
Again, ${\bf A}$ is independent of $t$. If the time series is defined as a linear projection 
$f = {\bf p}^T {\bf f} $ and ${\cal L} \left[ {\bf f} (t) \right] = {\bf M} {\bf f}$,
an equivalent analysis as above shows that ${\bf M}$ and ${\bf A}$ have identical spectra.
%The equivalent of Eq.~\ref{u_v_symm} now reads:
%\begin{eqnarray}
%	A _{k l} & = & \frac{1}{\sqrt{\lambda _l}} \int _{\Omega _W} d t u_l (t) 
%		{\cal L} \left[ u _k (t) \right] \nonumber
%\end{eqnarray}	
%where the asymmetric relation between the functions $u_k (t)$ associated with the manifold $M$,
%and the functions $v_k ( \xi )$ associated with the group $G$, is apparent.

Lie groups was originally developed to analyze symmetries in differential equations,
see e.g.,~\cite{Olver}. To make a direct connection between this theory and the analysis
above, assume that $f(t)$ is a solution to some linear differential equation,
i.e., $D _t f (t) = 0$ where $D_t$ is a linear operator. Assume further that $\Psi _{\xi}$ is a
symmetry group of the differential operator $D_t$, then $D_t f (t) = 0$ implies $D_t 
\left( \Psi _{\xi} f (t) \right) = 0$. If the transformation group $\Psi _{\xi}$
is used to construct the trajectory function, then the principal eigenfunctions 
$u _k (t)$ will also satisfy the same differential equation, $D_t u _k (t) = 0$. This can be 
seen by noting that the t-covariance kernel can be written explicitly as
\begin{eqnarray}
S \left( t , t ^{\prime} \right) & = & \int _{\Omega _{T-W}} \! \! \! \! 
	\! \! \! \! \! \!  d \xi f 
	\left( \Psi _{\xi} t \right) f \left( \Psi _{\xi} t ^{\prime} \right) .
\end{eqnarray}
Since $D _t S \left( t , t^{\prime} \right) = 0$, it follows from the eigenvalue problem in
Eq.~\ref{eigen_u_v} that $D_t u _k (t) = 0$, if $\lambda _k \neq 0$.

The analysis also shows how the global transformation group, used
in the construction of the trajectory function, affects the eigenmodes
via the infinitesimal generator. Let the vector field be defined by 
${\bf V} = \zeta (t) \partial _t$. The relation between the eigenmodes in
Eq.~\ref{u_general} and Eq.~\ref{u_standard} is then given by a variable
transformation, $\tau = \int ^t \frac{d s}{\zeta ( s )}$. The general
solutions is on the form
\begin{eqnarray}
	{\bf u} (t) & = & \exp \left\{ {\bf A} \int ^t \frac{d s}{ \zeta ( s ) } 
		\right\} {\bf u}_0 .
\label{u_solutions}
\end{eqnarray}
Eq.~\ref{u_solutions} reflects the fact that up to an isomorphism, there are only
two connected one-parameter Lie groups, ${\bf R}$ and $SO (2)$, corresponding
to real respective imaginary eigenvalues of ${\bf A}$, 
see~\cite{Olver} for details.

In the standard SSA, time translations are used to 
construct the trajectory function, corresponding to $\Psi (\xi , t ) = \xi + t$, 
${\bf V} = \partial _t$ and ${\bf u} (t) =  \exp \left\{ {\bf A} t \right\}{\bf u}_0 $. Using scale 
transformations, $\Psi (\xi , t ) = e^{ \xi } t$, corresponds to ${\bf V} = t \partial _t$
and ${\bf u} (t) =  \exp \left\{ {\bf A} \ln (t) \right\}{\bf u}_0 $, which contain
functions on the form $t^{\alpha }$ (scaling functions) and $\sin \left( \omega 
\ln (t) \right)$.

In fact, any smooth vector field generates a transformation group, which may 
only be locally defined. For example, consider $\zeta (t) = t ^{\alpha}$
for some constant $\alpha \in {\bf R} , \alpha \neq 1$. Using Eq.~\ref{orbit} and 
Eq.~\ref{u_solutions}, we then have:
\begin{eqnarray}
	{\bf V} & = & t ^{\alpha } \partial _t \nonumber \\
	\Psi \left( \xi , t \right) & = &  \left( \beta \xi + t^{\beta} 
		\right) ^{\frac{1}{\beta}} \nonumber \\
	{\bf u} (t) & = &  \exp \left\{ {\bf A} \beta ^{-1} t^{\beta } 
		\right\} {\bf u}_0 ,
\nonumber
\end{eqnarray}
where $\beta = 1 - \alpha $.

For completeness, we finish this paper by a further generalization of the SSA. 
Consider a smooth function $\psi : \Omega _W \times \Omega _{T-W} \rightarrow \Omega _T$, 
and construct the trajectory function as
$X (\xi , t) = f \left( \psi ( \xi , t) \right)$. We assume the mapping
to be of maximal rank, i.e., have a non-vanishing
Jacobian. The implicit function theorem then ensures that in the neighborhood of
each point on a curve defined by $\psi ( \xi , t ) = const.$, $\xi$
can be expressed as a function of $t$ and vice versa, i.e., the mapping 
$\psi$ defines an implicit one-dimensional
submanifold in ${\bf R}^2$. We assume
$\psi \left( 0 , t \right) = t$. For small $\xi ^{\prime}$, we then use the
smoothness of $\psi$ to write
\begin{eqnarray}
	\psi \left( \xi , \psi \left( \xi ^{\prime} , t \right) \right) & = &
	\psi \left( \xi + h(t) g ( \xi , t ) ( \xi ^{\prime} )^{\kappa} , t  \right)  \nonumber \\
	\psi \left( \xi ^{\prime} , t \right) & = & t + h (t)  ( \xi ^{\prime} )^{\kappa} , 
\label{gen_fnc}
\end{eqnarray}
which is valid to order $\kappa \geq 1$ in $\xi ^{\prime}$ for some functions $h$ and $g$. 
The derivative of the first relation in Eq.~\ref{gen_fnc} gives an explicit expression
for $g ( \xi , t )$:
\begin{eqnarray}
	g ( \xi , t ) & = & \frac{\partial _t \psi ( \xi , t)}{\partial _{\xi} 
		\psi ( \xi , t)}  
\nonumber
\end{eqnarray}
Note that if $h(t) g \left( \xi , t \right) = 1$ and $\kappa = 1$, then $\psi$ is a transformation
group according to Definition~\ref{trans}. Given a function $g ( \xi , t)$ we can also
find the global mapping $ \psi $ by solving the linear first order 
PDE (using for example the method of characteristics):
\begin{eqnarray}
	\partial _t \psi (\xi , t ) - g ( \xi , t) \partial _{\xi} \psi ( \xi , t) & = & 0 \nonumber \\
	\psi ( 0 , t) & = & t , 
\label{PDE}
\end{eqnarray}
corresponding to Eq.~\ref{orbit} above. Using Eq.~\ref{gen_fnc}
we derive the structural differential equations for the principal eigenfunctions
\begin{eqnarray}
	\partial _t u _k (t) & = & \sum _l A _{k l} (t) u _l (t) 
\label{u_more_general}
\end{eqnarray}
where
\begin{eqnarray}
	A _{k l} (t) & = & \sqrt{ \frac{\lambda _l}{\lambda _k}} \int _{\Omega _{T-W}} \! \! \! \! 
	\! \! \! \! \! \!   d \xi  g \left( \xi , t \right) v _k ( \xi ) 
	\partial _{\xi} v _l ( \xi ),
\nonumber
\end{eqnarray}  
which in this case dependent on $t$. The time dependence of ${\bf A}$ in Eq.~\ref{u_more_general} 
makes the equations hard to analyze in general. Again it  worth noting that an 
equivalent approach would be to use the relation
\begin{eqnarray}
	\partial _{\xi} \psi ( \xi , t ) \partial _{\xi}
f \left( \psi ( \xi , t ) \right)  = \partial _t \psi ( \xi , t ) \partial _t
f \left( \psi ( \xi , t ) \right),
\nonumber
\end{eqnarray}
which is valid for all smooth functions $f$ (especially $f ( x ) = x$ which gives Eq.~\ref{PDE}).

There are a number of different ways to further generalize SSA along the lines of this paper.
The perhaps most natural is to consider multi-dimensional transformation groups and/or
multi-dimensional data fields. This would require an extension of the SVD to
tensor decomposition. This is not straightforward~\cite{Kolda}, but generalized
versions of the SVD does exist~\cite{Lathauwer}.

Finally, the author would like to thank Steen Rasmussen, for support as well as valuable
discussions and perspectives. The author would also like to acknolege grant support from
U.S. departmet of Energy.

\end{document}